\documentclass[aps,prb,superscriptaddress,showpacs,twocolumn]{revtex4}
\usepackage{amsfonts}
\usepackage{amsmath}
\usepackage{amssymb}
\usepackage{graphicx}

\providecommand{\U}[1]{\protect\rule{.1in}{.1in}}
\begin{document}

\title{Anomalous Josephson Effect in magnetic Josephson junctions with
noncentrosymmetric superconductors}
\author{Huan Zhang}
\affiliation{Department of Physics, South University of Science
and Technology of China, Shenzhen 518055, P.R. China}
\author{Jun Wang}
\affiliation{Department of Physics, Southeast University, Nanjing
210096, China}
\author{Jun-Feng Liu}
\email{liujf@sustc.edu.cn}
\affiliation{Department of Physics, South University of Science and Technology of China,
Shenzhen 518055, P.R. China}

\begin{abstract}
We show that the two-band nature of noncentrosymmetric
superconductors leads naturally to an anomalous Josephson current
appearing at zero phase difference in a clean noncentrosymmetric
superconductor/ferromagnet/noncentrosymmetric superconductor
junction. The two-band nature provides two sets of Andreev bound
states which carry two supercurrents with different amplitudes.
When the magnetization direction of the ferromagnet is suitably
chosen, two supercurrents experience opposite phase shifts from
the conventional sinusoidal current-phase relation. Then the total
Josephson current results in a continuously tunable ground-state
phase difference by adjusting the ferromagnet parameters and the
triplet-singlet ratio of noncentrosymmetric superconductors. The
physics picture and analytical results are given on the basis of
the $s$+$p$ wave, while the numerical results are reported on both $s$+$p$ and
$d$+$p$ waves. For the $d$+$p$ wave, we find novel states in which the
supercurrents are totally carried by continuous propagating states
instead of discrete Andreev bound states. Instead of carrying
supercurrent, the Andreev bound states which here only appear
above the Fermi energy block the supercurrent flowing along the
opposite direction. These novel states advance the understaning of
the relation between Andreev bound states and the Josephson
current. And the ground-state phase difference serves as a tool to
determine the triplet-singlet ratio of noncentrosymmetric
superconductors.
\end{abstract}

\maketitle

\section{Introduction}

The noncentrosymmetric superconductor (NCS) has attracted much
attention for the coexistence of spin-singlet and spin-triplet
superconductivity and the possibility of topologically nontrivial
surface states\cite{brydon11,schnyder11,yada11,tanaka10,sato10,timm10,eschrig08}.
The unusual properties of NCS originate from the absence of
inversion symmetry from the crystal structure, which permits an
antisymmetric spin-orbit coupling (SOC) odd in electron momentum,
and leads to a chiral ground state. The SOC mixes the spin-singlet
(even-parity) component and the spin-triplet
(odd-parity) component in the superconducting pairing potential \cite%
{rashba01,kimura05,sugitani06}. The list of NCSs has grown to include dozens
of materials such as Li$_{2}$Pd$_{x}$Pt$_{3-x}$B \cite{togano04,togano05}, Y$%
_{2}$C$_{3}$ \cite{amano04}, and the heavy-fermion compounds CePt$_{3}$Si
\cite{bauer04}, CeRhSi$_{3}$ \cite{kimura05}, and CeIrSi$_{3}$ \cite%
{sugitani06}. These compounds have various crystal structures and hence
various forms of SOC and mixed triplet-singlet pair symmetry. The actual
superconducting pairing mechanics and pairing state symmetry realized in
these NCSs are still unclear.

In CePt$_3$Si and several other Ce-based NCSs, the spin-singlet and the
spin-triplet components are expected to appear in comparable magnitudes \cite%
{bauer04}, which may lead to more exotic effects. As a result, the NCS has
two effective superconducting gaps that are relevant to two sets of
spin-split Andreev bound states. The relative magnitude of the two parity
components determines the relative size of the two gaps and the main
property of a NCS. The question of how to determine the triplet-singlet
ratio has attracted many theoretical and experimental efforts \cite%
{borkje06,asano11,klam14,rahnavard14,tanaka07,fujimoto09,klam09,yuan06}. As
one of the main experimental methods in the field of superconductivity, the
Josephson effect has been widely studied in junctions based on NCSs to
investigate the characteristic triplet-singlet ratios. There were efforts
focusing on the investigations of the steps in the current-voltage
characteristics \cite{borkje06}, the low-temperature anomaly in the critical
current \cite{asano11}, and the transition from a $0$ junction to a $\pi/2$
junction \cite{klam14} in Josephson junctions with NCSs. The magnetic
Josephson junction with NCSs has also been studied \cite{rahnavard14} and
the authors focused on the high-order harmonics in the charge and spin
current-phase relations and the possibility of $0$-$\pi$ transitions.

On the other hand, much attention has also been paid to the anomalous
Josephson effect which means a $\varphi_0$-junction with arbitrary
ground-state phase difference $\varphi_0$ other than $0$ or $\pi$. Usually,
the supercurrent in a Josephson junction vanishes when the phase difference
between the two superconductors becomes zero and the current-phase relation
(CPR) is sinusoidal $I(\varphi)=I_{c}\sin(\varphi)$ in the tunnelling limit
\cite{golubov04}. Whereas an anomalous Josephson current $I_{a}$ flowing
even at zero phase difference ($\varphi=0$) has recently been predicted in
various types of Josephson junctions \cite%
{brydon08,krive1,feinberg,buzdin,martin09,liu10,liu102,liu11,liu14}. The
anomalous supercurrent is equal to a phase shift $\varphi_{0}$ in the
conventional CPR, i.e., $I(\varphi)=I_{c}\sin(\varphi-\varphi_{0})$. In
general, there are two prerequisites for the emergence of a $\varphi_0$%
-junction: (i) two sets of spin-split Andreev bound states (ABS)
with opposite phase shifts $\pm\varphi_{0}$ compared with the
conventional CPR, and (ii) different amplitudes of the
supercurrents carried by the two sets of ABS. Especially, the
Josephson junction formed on the surface of topological insulators
by the proximity effect can be considered as the limiting case
where only one set of ABS remain \cite{tanaka09}.

In Josephson junctions based on NCSs, the second prerequisite is
naturally met due to the two-band nature of superconductivity in
NCSs. The ABS are naturally spin-split into two sets and the two
gaps with different sizes ensure that the supercurrents carried by
two sets of ABS have different amplitudes. To meet the first
prerequisite, we introduce a ferromagnet into the Josephson
junction which brings phase shifts to ABS. In this study, we
investigate the anomalous Josephson effect in a noncentrosymmetric
superconductor/ferromagnet/noncentrosymmetric (NCS/F/NCS)
junction. It is shown that the ground-state phase difference
$\varphi_{0}$ is sensitive to the triplet-singlet ratio of NCSs.
Therefore, the anomalous Josephson effect serves as a mechanism to
determine the unknown triplet-singlet ratio of a NCS. The physics
picture and analytical results are given on the basis of the $s$+$p$ wave,
while the numerical results are reported on both $s$+$p$ and $d$+$p$ waves.
For $d$+$p$ wave, we find novel states in which the supercurrents are
totally carried by continuous propagating states instead of
discrete Andreev bound states.

The paper is organized as follows. In Sec. II we present the model
Hamiltonian and introduce the numerical method based on the lattice Green's
function to solve the CPR. In Sec. III we present the analytical results of
the normal incidence component in the case of s+p wave. The numerical
results and relevant discussion on three types of pair potentials s+p, d+p,
d+f will be given in Sec. IV. Finally, the conclusion will be given in Sec.
IV.

\section{Model and Numerical Methods}

\begin{figure}[tbp]
\begin{center}
\includegraphics[bb=15 16 312 155, width=3.405in]{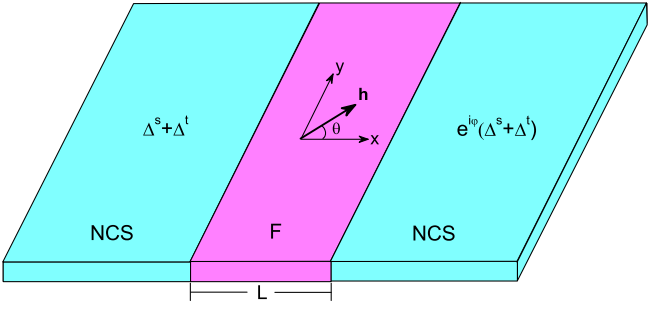}
\end{center}
\caption{(Color online) Schematic diagram of the NCS/F/NCS junction. The
ferromagnetic layer F has a finite width $L$. The exchange field $\mathbf{h}$
lies in the $x$-$y$ plane and makes an angle $\protect\theta$ with the $x$%
-axis.}
\label{model}
\end{figure}

We consider a two-dimensional NCS/F/NCS junction in the clean limit. A
schematic diagram of the junction under study is shown in Fig. \ref{model}.
The ferromagnetic layer F has a finite width $L$, and an exchange field $%
\mathbf{h}$ whose direction is in the $x$-$y$ plane and makes an angle $%
\theta$ with the $x$-axis.

The numerical method used to evaluate the supercurrent is the lattice Green's
function technique. We consider the junction in a square lattice with
the lattice constant $a$. The lattice lies in the $x$-$y$ plane. The
ferromagnetic layer F sandwiched by the two NCS electrodes locates in the
region $[0,L=Na]$ with $N$ is the number of columns. During the tunnelling
processes in the $x$-direction, the transverse momentum $k_{y}$ is assumed
to be conserved. In this context, the Hamiltonian of the hybrid junction
reads%
\begin{align}
H& =\sum\limits_{l\sigma ,k_{y}}\left[ \varepsilon _{l,k_{y}}-\mu \right]
c_{l\sigma ,k_{y}}^{\dagger }c_{l\sigma ,k_{y}}-t\sum\limits_{l\sigma
,k_{y}}c_{l\sigma ,k_{y}}^{\dagger }c_{l\pm 1\sigma ,k_{y}}  \notag \\
& \text{ \ \ }+\sum\limits_{l,l^{\prime },k_{y}}\sum\limits_{\sigma \sigma
^{\prime }}\left( h_{l\sigma ,l^{\prime }\sigma ^{\prime }}+\lambda
_{l\sigma ,l^{\prime }\sigma ^{\prime }}\right) c_{l\sigma ,k_{y}}^{\dagger
}c_{l^{\prime }\sigma ^{\prime },k_{y}}  \notag \\
& \text{ \ \ }-\sum\limits_{l,l^{\prime },k_{y}}\sum\limits_{\sigma \sigma
^{\prime }}\left[ \Delta _{l\sigma ,l^{\prime }\sigma ^{\prime }}c_{l\sigma
,k_{y}}^{\dagger }c_{l^{\prime }\sigma ^{\prime },-k_{y}}^{\dagger }+h.c.%
\right] ,  \label{HAMILTONIAN}
\end{align}%
where $c_{l\sigma ,k_{y}}^{\dagger }$($c_{l\sigma ,k_{y}}$) is the creation
(annihilation) operator of an electron in column $l$ with spin $\sigma $ (= $%
\uparrow $ or $\downarrow $) and transverse momentum $k_{y}$. The on-site
energy $\varepsilon _{l,k_{y}}$ has the form $-2t\cos K_{y}$ with $%
K_{y}=k_{y}a$. The Fermi energy $\mu$ and the nearest-neighbor hopping
integral $t$ are assumed to be the same in the whole junction. The hopping
coefficients of the exchange interaction $h_{l\sigma ,l^{\prime }\sigma
^{\prime }}$ and the Rashba SOC $\lambda _{l\sigma ,l^{\prime }\sigma
^{\prime }}$ are given by the matrices in the spin space
\begin{eqnarray}
\hat{h}_{ll^{\prime }} &=&\mathbf{h}\cdot \boldsymbol{\sigma }\delta
_{l,l^{\prime }}  \notag  \label{EXCH} \\
\hat{\lambda}_{ll^{\prime }} &=&\lambda (\sigma _{x}\sin K_{y}\delta
_{l,l^{\prime }}\mp i\sigma _{y}\delta _{l\mp 1,l^{\prime }}/2)  \label{RSOC}
\end{eqnarray}%
where $\mathbf{h}$ and $\lambda $ denote the exchange field and the Rashba
strength respectively. The pair potential coefficient $\Delta _{l\sigma
,l^{\prime }\sigma ^{\prime }}$ reads\cite{Asano01,Asano06}%
\begin{widetext}
\begin{equation}
\hat{\Delta}_{ll^{\prime }}=\left\{
\begin{array}{l}
\Delta \left[ \left( q+\bar{q}\sin K_{y}\sigma _{x}\right) \delta
_{l,l^{\prime }}\mp \frac{i}{2}\bar{q}\sigma _{y}\delta _{l\mp 1,l^{\prime }}%
\right] i\sigma _{y}\text{,} \\
\Delta \left\{ \left[ \pm \frac{i}{2}q\sin K_{y}\pm
i\bar{q}(1-\cos K_{y})\sigma _{x}+\bar{q}\sin K_{y}\sigma
_{y}\right] \delta _{l\mp 1,l^{\prime }}-2\bar{q}\sin K_{y}\sigma
_{y}\delta _{l,l^{\prime }}\right\} i\sigma _{y}\text{,} \\
\Delta \left[ \left( 2\cos K_{y}\delta _{l,l^{\prime
}}-\delta _{l\mp
1,l^{\prime }}\right) \left( q+\bar{q}\sin K_{y}\sigma _{x}\right) +\bar{q}%
(\mp i\cos K_{y}\delta _{l\mp 1,l^{\prime }}\pm \frac{i}{2}\delta _{l\mp
2,l^{\prime }})\sigma _{y}\right] i\sigma _{y}\text{,}%
\end{array}%
\right. \left.
\begin{array}{l}
s+p\text{-wave} \\
d_{xy}+p\text{-wave} \\ \label{pair}
d_{x^{2}-y^{2}}+p\text{-wave}%
\end{array}%
\right.
\end{equation}
\end{widetext}
where $\Delta $ is the BCS gap function which takes $\Delta _{0}$ at zero
temperature and vanishes at critical temperature $T_{c}$. For simplicity,
the pair potentials in the two NCS leads are set to be equal in amplitude. $%
\sigma _{i}$ with $i=x$, $y$, $z$ are the Pauli matrices. The parameter $q$ (%
$\bar{q}=1-q$) characterizes the percentage of spin-singlet(triplet)
component which reads $q_{L}\left( \bar{q}_{L}\right) $ and $q_{R}\left(
\bar{q}_{R}\right) $ in the left and right NCS respectively. In Eq. $\left( %
\ref{pair}\right) $, we also omit the phase difference between the two NCSs
for simplicity. The phase difference is set to be $\varphi $. Note that the
exchange field exists only in the F layer while the Rashba SOC and the pair
potential only exist in the two NCSs.

In the F region ($0\ll l\ll N)$, the charge operator in column $l$ with
momentum $k_{y}$ is defined as%
\begin{equation}
\hat{\rho}_{l,k_{y}}=e\tilde{c}_{l,k_{y}}^{\dagger }{\sigma }_{0}\tilde{c}%
_{l,k_{y}},  \label{CHOP}
\end{equation}%
where $\tilde{c}_{l,k_{y}}=\left[ c_{l\uparrow ,k_{y}}\left( \tilde{t}%
\right) ,c_{l\downarrow ,k_{y}}\left( \tilde{t}\right) \right] ^{T}$, ${%
\sigma }_{0}$ is the unite matrix, and $\tilde{t}$ is the time. By using the
Heisenberger equation $i\hbar \partial _{\tilde{t}}\hat{\rho}=\left[ \hat{%
\rho},H\right] $, the operator of supercurrent is found. Then we can
construct the Green's function to calculate the supercurrent $I$ through
column $l$ as follows \cite{JWang10}%
\begin{equation}
I=\frac{1}{2\pi }\int \text{Tr}\left[ \check{t}^{\dag }\check{e}G^{<}\left(
l,l-1\right) -\check{e}\check{t}G^{<}\left( l-1,l\right) \right] dK_{y}
\label{SUPERI}
\end{equation}%
where $\check{t}=-t\tau _{3}\otimes {\sigma }_{0}$ and $\check{e}=-e\tau
_{3}\otimes {\sigma }_{0}$ denote the hopping matrix and the charge matrix
respectively. $\tau _{3}$ is the Pauli matrix in Nambu space and $e$ $\left(
>0\right) $ is the unit charge. In equilibrium, the lesser-than Green's
function $G^{<}$ equals

\begin{equation}
G^{<}=\int \frac{dE}{2\pi \hbar }f\left( E\right) \left[ G^{a}-G^{r}\right]
\label{EQUI}
\end{equation}%
where $f\left( E\right) $ is the Fermi-Dirac distribution function. The
retarded (advanced) Green's function $G^{r}\left( G^{a}\right) $ can be
numerically calculated by the recursive method.

Besides the supercurrent, the ABS spectra can be also calculated  numerically.
It is known that the ABS results in the peaks of particle
density within the superconducting gap. Therefore, by searching the peaks of
particle density in column $l$ $\left( N\geqslant l\geqslant 1\right) $

\begin{equation}
\rho _{l}=-\frac{1}{\pi }Im\left[ \text{Tr}\left\{ G^{r}\left( l,l\right)
\right\} \right]  \label{SABS}
\end{equation}%
at a given phase difference $\varphi $, the energies of ABS can be located.
Then we scan $\varphi $ and obtain the ABS spectrum which is useful for
understanding the behavior of supercurrent.

\section{Analytical Results}

Before we discuss our numerical results, we present the analytical results
of the normal incidence component in the $s$+$p$ wave case, which is very
helpful to understand how the junction under consideration becomes a $%
\varphi _{0}$-junction. We start with the Bogoliubov-de Gennes (BdG)
Hamiltonian for a NCS in the momentum space%
\begin{equation}
H=\allowbreak \left(
\begin{array}{cc}
\varepsilon _{\mathbf{k}}+\lambda \mathbf{l}_{\mathbf{k}}\cdot \mathbf{%
\sigma } & \Delta (\mathbf{k}) \\
\Delta ^{\dag }(\mathbf{k}) & -\varepsilon _{\mathbf{k}}+\lambda \mathbf{l}_{%
\mathbf{k}}\cdot \mathbf{\sigma }^{\ast }%
\end{array}%
\right)
\end{equation}%
Here, $\varepsilon _{\mathbf{k}}=\frac{\hbar ^{2}(k_{x}^{2}+k_{y}^{2})}{2m}%
-\mu =\frac{\hbar ^{2}k^{2}}{2m}-\mu $ is the spin-independent part of band
dispersion with $\mu $ the chemical potential, and $\mathbf{\sigma }$ is the
vector of Pauli matrices. $\lambda \mathbf{l}_{\mathbf{k}}\cdot \mathbf{%
\sigma }$ is the antisymmetric SOC with $\mathbf{l}_{\mathbf{k}}=k_{y}%
\widehat{\mathbf{x}}-k_{x}\widehat{\mathbf{y}}$ and $\lambda $ the SOC
strength. The superconducting gap function is $\Delta (\mathbf{k})=f(\mathbf{%
k})(\Delta _{s}+\Delta _{t}\mathbf{d}_{\mathbf{k}}\cdot \mathbf{\sigma }%
)i\sigma _{y}$, where $\Delta _{s}=\Delta _{0}q$ and $\Delta _{t}=\Delta
_{0}(1-q)$ are spin-singlet and spin-triplet superconducting gaps
respectively with $\Delta _{0}=\Delta _{s}+\Delta _{t}$ and $q$ turns
between purely spin-triplet ($q=0$) and purely spin-singlet ($q=1$)
pairings. We assume $\Delta _{s}$, $\Delta _{t}$, and $\Delta _{0}$ are
positive constants, and the orbital-angular-momentum pairing state is
described by the structure factor $f(\mathbf{k})$. In the case of s+p wave, $%
f(\mathbf{k})=1$. The spin-triplet pairing vector is aligned with the
polarization vector of the SOC $\mathbf{d}_{\mathbf{k}}=\mathbf{l}_{\mathbf{k%
}}/k_{F}$ where $k_{F}$ is the Fermi wave vector and taken as the unit of
the wave vector. When the SOC splitting is much less than the chemical
potential, we can use the Andreev approximation $k_{F}^{\pm }\approx k_{F}=1$
with $k_{F}^{\pm }$ the spin-split Fermi wave vectors.

To diagonalize the kinetic term, it is convenient to express the Hamiltonian
in the so-called helicity basis. We introduce the following spin rotation
transformation
\begin{equation}
R=\allowbreak \left(
\begin{array}{cc}
U & 0 \\
0 & U^{\ast }%
\end{array}%
\right) ,U=\frac{1}{\sqrt{2}}\left(
\begin{array}{cc}
1 & -ie^{-i\phi } \\
1 & ie^{-i\phi }%
\end{array}%
\right),
\end{equation}%
where $\phi =\arctan (k_{y}/k_{x})$ is the incident angle of quasiparticles.
Under the rotation $R$, the normal part of the Hamiltonian is diagonalized
and the superconducting gap function is transformed to

\begin{equation}
U\Delta (\mathbf{k})\left( U^{\ast }\right) ^{-1}=\allowbreak \tilde{f}(%
\mathbf{k})\left(
\begin{array}{cc}
0 & \Delta _{t}-\Delta _{s} \\
\Delta _{t}+\Delta _{s} & 0%
\end{array}%
\right)
\end{equation}%
with $\allowbreak \tilde{f}(\mathbf{k})=-ie^{-i\phi }\allowbreak f(\mathbf{k}%
)$. The factor $-ie^{-i\phi }\allowbreak $ is the same for the left and
right NCSs and thus has no net effect on the Josephson effect in the first
harmonic approximation where the normal reflection is absent. The
Hamiltonian of a NCS in the hilicity basis reads%
\begin{equation}
H=\allowbreak \left(
\begin{array}{cccc}
\varepsilon _{\mathbf{k}}-\lambda k & 0 & 0 & \Delta _{-}\allowbreak \tilde{f%
}(\mathbf{k}) \\
0 & \varepsilon _{\mathbf{k}}+\lambda k & \Delta _{+}\allowbreak \tilde{f}(%
\mathbf{k}) & 0 \\
0 & \Delta _{+}\allowbreak \tilde{f}(\mathbf{k}) & -\varepsilon _{\mathbf{k}%
}-\lambda k & 0 \\
\Delta _{-}\allowbreak \tilde{f}(\mathbf{k}) & 0 & 0 & -\varepsilon _{%
\mathbf{k}}+\lambda k%
\end{array}%
\right)  \label{h44}
\end{equation}%
where $\Delta _{\pm }=\Delta _{t}\pm \Delta _{s}$. It is clear that the
Hamiltonian shows a two-band nature that there are two bands with different
superconducting gaps $\left\vert \Delta _{\pm }\right\vert $ are uncoupled
in the helicity basis. One band is for the Cooper pair made of spin-up
electron and spin-down hole with gap $\left\vert \Delta _{-}\right\vert $,
the other band is for the pair of spin-down electron and spin-up hole with
gap $\Delta _{+}$ with respect to the helicity basis. Since the critical
Josephson current is linear in the gap, the two bands provide two
supercurrents with different amplitudes. Thus the second prerequisite for
the emergence of a $\varphi _{0}$-junction is naturally reached.

The first prerequisite for a $\varphi _{0}$-junction is easy to meet by
utilizing a ferromagnetic interlayer. The middle F layer can bring opposite
phase shifts to the two supercurrents provided by the two bands because the
spin-triplet component of the gap function is also spin-opposite pairing in
the helicity basis. For simplicity, we consider only the normal incidence
component with $\phi =0$ which is dominant in the Josephson current for the
case of orbital $s+p$ wave. We consider a F interlayer with the exchange
filed strength $h$ and the width $L$. The magnetization direction is chosen
to be aligned with the polarization direction of the SOC, i.e., the $y$%
-direction. In this situation, the two helical bands keep uncoupled from
each other and the Hamiltonian of the whole NCS/F/NCS junction for the two
bands is respectively%
\begin{equation}
H_{\sigma }=\allowbreak \left(
\begin{array}{cc}
\varepsilon _{\mathbf{k}}-\sigma \lambda k+\sigma \widetilde{h}(x) & \Delta
_{\overline{\sigma }}(x)\allowbreak \tilde{f}(\mathbf{k}) \\
\Delta _{\overline{\sigma }}(x)\allowbreak \tilde{f}(\mathbf{k}) &
-\varepsilon _{\mathbf{k}}+\sigma \lambda k+\sigma \widetilde{h}(x)%
\end{array}%
\right)  \label{h22}
\end{equation}%
where the helicity index $\sigma =1$ is for the pair of spin-up electron and
spin-down hole while $\sigma =-1$ is for the pair of spin-down electron and
spin-up hole with respect to the helicity basis, $\widetilde{h}%
(x)=sgn(k_{x})h(x)$ with $h(x)=h[\Theta (x)-\Theta (x-L)]$, $\Delta _{%
\overline{\sigma }}=$ $\Delta _{\mp }$ when $\sigma =\pm 1$ and $\Delta _{%
\overline{\sigma }}(x)=\Delta _{\mp }[\Theta (-x)e^{i\varphi /2}+\Theta
(x-L)e^{-i\varphi /2}]$ with $\varphi $ the macroscopic phase difference of
the two NCSs. The F layer is expected to bring phase shifts $\sigma \eta $
to the set of ABS with index $\sigma $ and the corresponding supercurrent.
In the dimensionless units, the phase shift is approximately \cite{tanaka00}
$\eta =(k_{-}-k_{+})L\approx hL$ with the wave vectors for spin-up or
spin-down $k_{\pm }=\sqrt{k^{2}\mp h}\approx k\mp h/2$. Note that the
right-going and left-going Cooper pairs with the same helicity index $\sigma
$ actually have opposite real spin with respect to the $y$-direction. It is
interesting that these two pairs have the same phase shift $\sigma \eta $
induced by the F layer \cite{tanaka00}, and correspond to the same gap $%
\Delta _{\overline{\sigma }}$ at the same time.

After the two prerequisites are reached, we come to discuss the supercurrent
carried by the two bands in the NCS/F/NCS junction by assuming that the ratio of
the supercurrent to the gap is a constant for the two bands. In the first
harmonic approximation, we have two supercurrents%
\begin{equation}
I_{\sigma }=\beta \left\vert \Delta _{t}-\sigma \Delta _{s}\right\vert \sin
(\varphi -\sigma \eta ).
\end{equation}%
The total Josephson current is the sum $I=I_{\sigma =1}+I_{\sigma =-1}$. It
is noticeable that the Josephson current depends on only $\left\vert \Delta
_{t}-\sigma \Delta _{s}\right\vert $ and it does not matter whether $\Delta
_{t}>\Delta _{s}$ or $\Delta _{t}<\Delta _{s}$ in the first harmonic
approximation. We refer to the bigger (smaller) one of $\Delta _{s}$ and $%
\Delta _{t}$ as $\Delta _{1}$ ($\Delta _{2}$). Then the total Josephson
current can be written as
\begin{eqnarray}
I &=&2\beta \left( \Delta _{1}\sin \varphi \cos \eta +\Delta _{2}\cos
\varphi \sin \eta \right)  \notag \\
&=&2\beta \sin (\varphi -\varphi _{0})
\end{eqnarray}%
with
\begin{eqnarray}
\sin \varphi _{0} &=&\frac{-\Delta _{2}\sin \eta }{\sqrt{\Delta _{1}^{2}\cos
^{2}\eta +\Delta _{2}^{2}\sin ^{2}\eta }},  \notag \\
\cos \varphi _{0} &=&\frac{\Delta _{1}\cos \eta }{\sqrt{\Delta _{1}^{2}\cos
^{2}\eta +\Delta _{2}^{2}\sin ^{2}\eta }}.  \label{phi0}
\end{eqnarray}%
It is seen that the anomalous ground-state phase difference $\varphi _{0}$
depends not only on the F layer induced phase shift $\eta =hL$ but also the
triplet-singlet ratio of the two NCSs. Thus the arbitrary $\varphi _{0}$%
-junction can be obtained by tuning the ferromagnet parameters and the
triplet-singlet ratio. We can also determine the triplet-singlet ratio of
NCSs by detecting the ground-state phase difference of such a NCS/F/NCS
junction.

\begin{figure*}[tbp]
\begin{center}
\includegraphics[bb=16 15 313 227, width=6.5in]{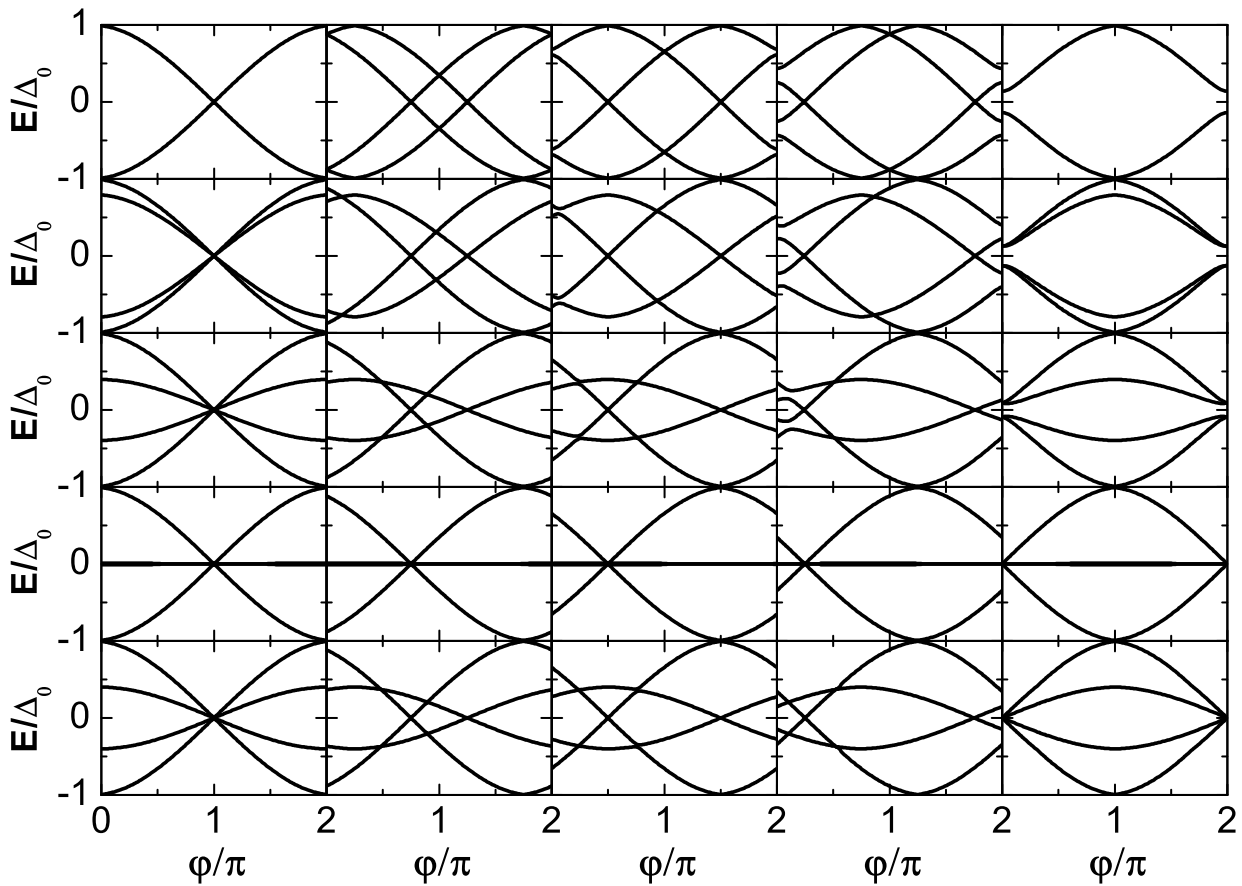}
\end{center}
\caption{Andreev bound states at normal incidence $k_y=0$ in a NCS/F/NCS
junction for $s$+$p$ wave. The exchange field $\mathbf{h}$ in the F layer is
aligned with the $y$-direction. From left to right, the exchange field
induced phase shift varies from $0$ to $\protect\pi$ with a step $\protect\pi%
/4$ by adjusting the exchange field strength $h$; from top to bottom, $%
q=0,0.1,0.3,0.5,0.7 $. Other parameters are: $\protect\mu=-2t$, $%
\Delta_0=0.01\protect\mu$, $\protect\lambda=0.1\protect\mu$.}
\label{abs}
\end{figure*}

In the above discussion, we assume that the left and right NCSs have the
same singlet percentage $q$. If the two NCSs have opposite $q$ parameters,
\textit{i.e.}, $q_{L}=1-q_{R}$ with $q_{L(R)}$ the $q$ parameter of left
(right) NCS, the situation is a little different. For the $\Delta _{-}$
band, the exchange of $\Delta _{s}$ and $\Delta _{t}$ changes the sign of
the gap function as shown in Eq. (\ref{h44}). Therefore, there will be
equivalently an additional phase difference $\pi $ between the left and
right NCSs. For the $\Delta _{+}$ band, the situation keeps unchanged. Then
the total Josephson current is
\begin{eqnarray}
I &=&-I_{\sigma =1}+I_{\sigma =-1}  \notag \\
&=&\beta \left[ \Delta _{+}\sin (\varphi +\eta )-\left\vert \Delta
_{-}\right\vert \sin (\varphi -\eta )\right]  \notag \\
&=&2\beta \left( \Delta _{2}\sin \varphi \cos \eta +\Delta _{1}\cos \varphi
\sin \eta \right)  \notag \\
&=&2\beta \sin (\varphi -\varphi _{0}^{\prime })
\end{eqnarray}%
with%
\begin{eqnarray}
\sin \varphi _{0}^{\prime } &=&\frac{-\Delta _{1}\sin \eta }{\sqrt{\Delta
_{2}^{2}\cos ^{2}\eta +\Delta _{1}^{2}\sin ^{2}\eta }},  \notag \\
\cos \varphi _{0}^{\prime } &=&\frac{\Delta _{2}\cos \eta }{\sqrt{\Delta
_{2}^{2}\cos ^{2}\eta +\Delta _{1}^{2}\sin ^{2}\eta }}.
\end{eqnarray}%
For the special case of $q_{L}=1$ and $q_{R}=0$, the result of a $\pi /2$
junction for the triplet-ferromagnet-singlet Josephson junction is recovered
as the same as in Ref. \onlinecite{brydon13}.

\section{Numerical Results and Discussion}

\subsection{s+p wave}

\begin{figure*}[tbp]
\begin{center}
\includegraphics[bb=17 15 313 214, width=5.815in]{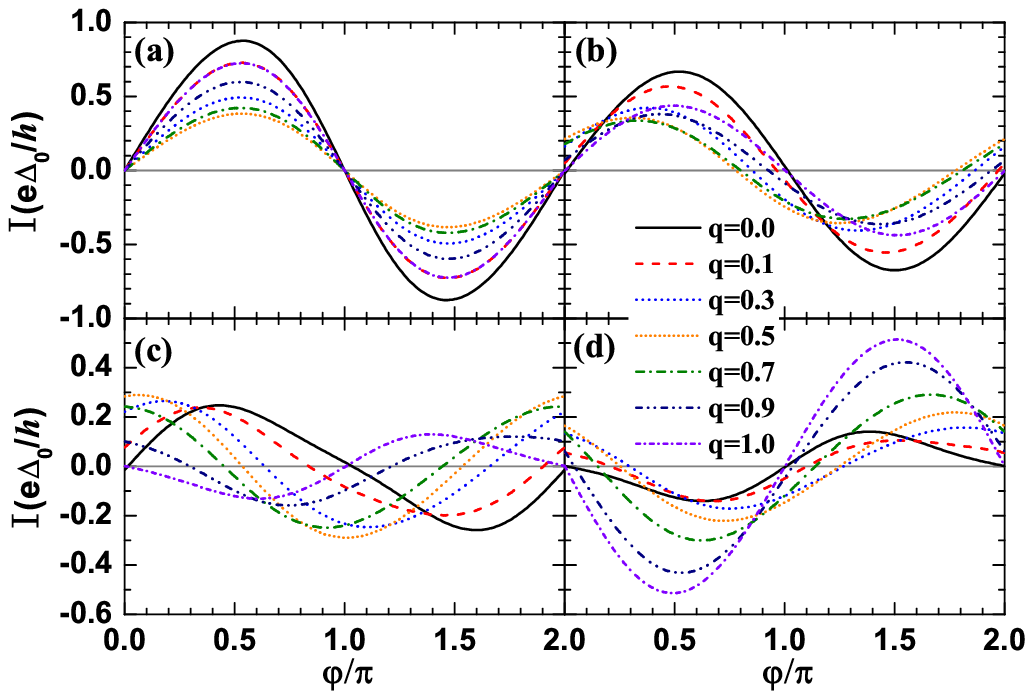}
\end{center}
\caption{(Color online) CPR of total Josephson current for the $s$+$p$ wave with
various phase shift $\protect\eta$: (a) $0$, (b) $\protect\pi/4$, (c) $%
\protect\pi/2$, (d) $3\protect\pi/4$. In each panel, the results for varying
values of $q$ from $0$ to $1$ are plotted. The temperature $T=0.9T_c$. Other
parameters are the same as those in Fig. \protect\ref{abs}.}
\label{sp}
\end{figure*}

For the $s$+$p$ wave case, the momentum dependence of the gap function $f(\mathbf{k})=1$.
The magnetization direction of the F layer is chosen to be aligned with the $%
y$-direction because the contribution from normal incidence $k_{y}=0$ is
dominant. In Fig. \ref{abs}, we show the numerically solved ABS at normal
incidence in the NCS/F/NCS junction. With increasing ferromagnet induced
phase shift $\eta $ (from left to right in Fig. \ref{abs}), two sets of ABS
with different helicity index $\sigma $ depart from each other more heavily.
The set of ABS with index $\sigma $ experience a phase shift $\sigma \eta $
from the original degenerate ABS. With increasing singlet percentage $q$,
the energy span of the set of ABS with helicity index $\sigma =1$ shrinks
due to the decreasing minor gap $\Delta _{\overline{\sigma }}=\Delta
_{-}=\Delta _{t}-\Delta _{s}$. That means the corresponding supercurrent
with $\sigma =1$ has a smaller amplitude. The special case of $q=0.5$ is
noticeable because the minor gap closes completely. Only the ABS with index $%
\sigma =-1$ is the remaining supercurrent-carried ABS. Then a $\varphi _{0}$%
-junction is easily realized by taking $\eta =-\varphi _{0}$, which agrees
well with Eq. (\ref{phi0}). The evolution of ABS with varying $\eta $ and $q$
shown in Fig. \ref{abs} is fully consistent with the above analytical
results.

It is also noteworthy that the ABS is almost the same for $q=0.3$ and $q=0.7$
as discussed in the analytical results. Because of the presence of weak
normal reflections at NCS/F interfaces in the numerical results, there opens
a small gap at some ABS crossing points due to the coupling of two ABS bands
with the same spin and the opposite travelling direction (thus with the
opposite helicity index $\sigma$). What is interesting is that the
anti-crossing effect for the case of $\Delta _{t}>\Delta _{s}$ is much
weaker than that for $\Delta _{s}>\Delta _{t}$. We can understand this
effect easily in two limit cases $q=0$ and $q=1$. For the coupled two bands
with the opposite $k_x$ and the opposite $\sigma$, Eq. (\ref{h22}) shows
that $\Delta _{t}$ changes sign while $\Delta _{s}$ does not. That is the
essential difference between the singlet and triplet pairing. Thus, the
normal reflection induced coupling results in an anti-crossing gap opening
for the triplet-dominant case but no obvious effect for the singlet-dominant
case at ABS crossing points. The anti-crossing effect gives rise to
high-order harmonics of the CPR, which is sensitive to the triplet-singlet
ratio as reported in Ref. \onlinecite{rahnavard14}. In this paper, we focus
on the first harmonic of the CPR.

\begin{figure*}[tbp]
\begin{center}
\includegraphics[bb=66 61 567 385, width=5.815in]{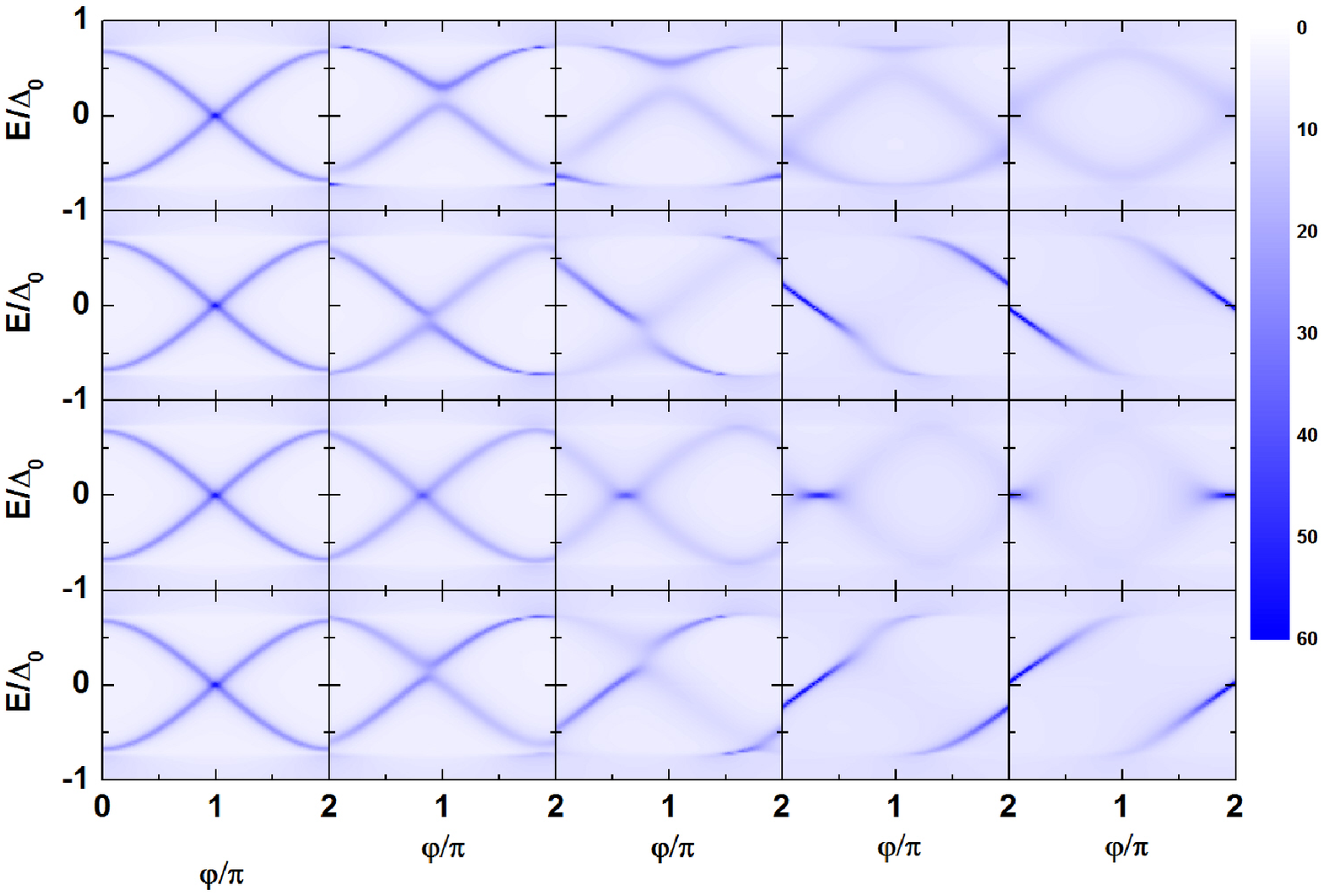}
\end{center}
\caption{(Color online) Particle density in the F layer at incidence angle $%
\protect\phi=\protect\pi/4$ for d+p wave. The singlet percentage
is fixed to $q=0.56$ to close the minor gap $\Delta_-$. From left
to right, the exchange
field induced phase shift $\protect\eta=0$, $\protect\pi/4$, $\protect\pi/2$%
, $3\protect\pi/4$, $\protect\pi$; from top to bottom, the angle
of the
magnetization direction $\protect\theta=0$, $\protect\pi/4$, $\protect\pi/2$%
, $3\protect\pi/4$, respectively. Other parameters are the same as
those in Fig. \protect\ref{abs}.} \label{absdp}
\end{figure*}

The CPR of the total Josephson current is shown in Fig. \ref{sp}. The
temperature is taken as $T=0.9T_c$ so that only the first harmonic is
remaining. When the F layer is absent, the CPR shows a normal $0$-junction
for all the values of $q$ as shown in Fig. \ref{sp} (a). While the critical
current decreases with increasing $q$ firstly till $q=0.5$ and then
increases when $q$ increases from $0.5$ to $1$. The change of the critical
current coincides well with the change of the minor gap in Fig. \ref{abs}
because the contribution from normal incidence $k_y=0$ is dominant. Although
the ABS are almost the same for $q$ and $1-q$ in the case of normal
incidence, the total Josephson current is not exactly the same for $q$ and $%
1-q$ because of the contribution from inclined incidence. For oblique
incidences, the spin-splitting of the Fermi surface is enhanced and will
modify the magnitude of triplet pairing.

When the phase shift $\eta =hL=\pi /4$, the CPR is shown in Fig.
\ref{sp} (b) for various $q$. For $q$ from $0$ to $1$, the
ground-state phase difference $\varphi _{0}$ reduces firstly to
the minimum $-\pi /4$ at $q=0.5$ and then goes back up to $0$. The
critical current also reaches its minimum at $q=0.5$. All these
features are qualitatively consistent with the ABS in Fig.
\ref{abs} and Eq. (\ref{phi0}). Unfortunately,  we cannot
distinguish the triplet-dominant pairing from the singlet-dominant
pairing just by the ground-state phase difference. However, such is not
the case when the phase shift $\eta =\pi /2$. It is shown in Fig.
\ref{sp} (c) that the ground-state phase difference of the CPR
decreases monotonically from $0$ to $-\pi $ when $q$ goes up from
$0$ to $1$. It is important that we can determine the
triplet-singlet ratio just by the ground-state phase difference of
the CPR. This special feature of $\eta =\pi /2$ can be understood
by considering three limit cases $q=0,0.5,1$. The case of $q=0.5$
is simple. The gap with $\sigma =1$ closes and the remaining set
of ABS with $\sigma =-1$ experience a phase shift $\sigma \eta
=-\pi /2$. Obviously we obtain a $-\pi /2$ junction. For the cases
of pure triplet or pure singlet pairing ($q=0$, or $1$), the
supercurrents carried by two sets of ABS have the opposite phase
shift $\pm \pi /2$ and the same amplitude. Then the two
supercurrents cannel each other out. Thus the contribution to
supercurrent from large incidence angles is dominant instead of
from normal incidence. For large incidence angle, the
magnetization direction ($y$-direction) is not aligned with the
spin direction of helicity basis any more. The singlet pairing is
always opposite-spin pairing (independent of the direction of the
spin quantization axis) while the triplet pairing is equal-spin
pairing in the spin quantization axes perpendicular to the
helicity basis. It is well-known that the F layer cannot bring a
phase shift for the equal-spin pairing. And when $k_{y}\neq 0$,
the wave vector difference $k_{x-}-k_{x+}>h$ for $q=1$. So the
larger the incident angle becomes, the less (more) the absolute
value of the phase shift is than $\pi /2$ for $q=0$
($1$). For pure pairings ($q=0$, or $1$), the CPR should be either a $0$%
-junction or a $\pi $-junction because of the presence of two supercurrents
with the opposite phase shift and the same amplitude. Therefore the CPR
tends to become a $0$-junction for $q=0$ while a $\pi $-junction for $q=1$.
When $q$ goes up from $0$ to $1$, the CPR naturally exhibits a $0$-$\pi $
transition with smoothly changed ground-state phase difference. When $\eta
=3\pi /4$, the ground-state phase difference varies between $-\pi $ and $%
-3\pi /4$, which is similar to the case of $\eta =\pi /4$ and also
consistent with the ABS in Fig. \ref{abs} and Eq. (\ref{phi0}). The
difference is that the critical current for $q>0.5$ is now larger than that
for $q<0.5$ because of the contribution from inclined incidence.

\subsection{$d$+$p$ wave}

For the $d$+$p$ wave NCS, the momentum dependence of the gap function $f(\mathbf{k}%
)=k_{x}k_{y}/k_{F}^{2}$. The dominant contributions come from the
two components with the incident angle $\phi=\pm \pi/4$. We cannot
choose a single magnetization direction which is aligned with both
spin polarization directions of these two components. That is to
say, the coupling of two helical bands is unavoidable for at least
one of the two components. For simplicity, we began with the
special case where only one gap survives and the other is
closed. The numerical results show that this can occur at $q=0.56$ when $%
\phi=\pm \pi/4$. The singlet percentage $q$ deviates from $0.5$
due to the lattice model and the spin-splitting of Fermi surface.
In Fig. \ref{absdp}, the particle density in the F layer at
incidence angle $\phi=\pi/4$ shows the evolution of ABS with
varied strength and direction of the exchange field. Because the
minor gap is closing, there are only two ABS remaining. The
right-going (left-going) ABS consists of a right-going
(left-going) electron and a left-going (right-going) hole. It is
noticeable that two ABS have different eigen spin direction
because $k_y$ is fixed and finite.

When $\theta =0$, the magnetization direction makes an angle of
$\pi /4$ with both spin-down directions ($\sigma =-1$ with respect
to the helical basis) of two ABS. On the one hand, the F layer
precesses the spin of electron and hole by an angle $\eta \approx
hL/\cos \phi $. Some spin-down particles are flipped to spin-up
and enter into the NCSs without Andreev reflection because the
spin-up gap is closed. The remaining spin-down particles proceed
to finish the cycle of two Andreev reflections to form a ABS. So
the particle density of two ABS shrinks with increasing $\eta $.
On the other hand, the F layer also brings a phase shift to the
two ABS. The phase shift is opposite for right-going and
left-going ABS. This is opposite to the previous case of the $s$+$p$ wave
where the phase shift is the same for right-going and left-going
ABS. In that case, the magnetization direction along the $y$-axis
is parallel to the spin-down direction of left-going ABS but
antiparallel to the spin-down direction of right-going ABS. When
$\theta =\pi /2$, the situation is similar. But the phase shift is
now the same for two ABS because the $y$-axis makes an angle of
$3\pi /4$ with the spin-down direction of right-going ABS while
$\pi /4$ with that of left-going ABS.

\begin{figure}[tbp]
\begin{center}
\includegraphics[bb=18 16 299 210, width=3.415in]{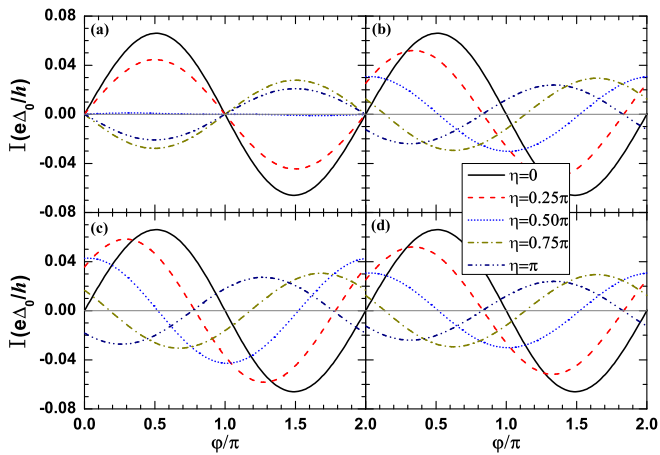}
\end{center}
\caption{(Color online) The CPR of total Josephson current for d+p
wave with fixed $q=0.56$ and various $\protect\theta$: (a) $0$,
(b) $\protect\pi/4$, (c) $\protect\pi/2$, (d) $3\protect\pi/4$. In
each panel, the results for varying values of $\protect\eta$ from
$0$ to $\protect\pi$ are plotted. Other parameters are the same as
those in Fig. \protect\ref{sp}.} \label{dp1}
\end{figure}

When $\theta =\pi /4$ and $3\pi /4$, the situation is particularly
interesting. Now the magnetization direction is parallel (or
antiparallel) to the spin-down direction of one ABS but
perpendicular to that of the other ABS. So one ABS experiences
only a phase shift while the other not only experiences a phase
shift but also shrinks a little. It is interesting that the shrinking ABS
disappears totally at $\eta =\pi $, which means that the spin-down
particles are flipped totally to spin-up and propagate into NCSs
as a continuous propagating state. For example, when $\theta =\pi
/4$ and $\eta =\pi $, the supercurrent is totally carried by
continuous propagating states instead of discrete ABS for the
phase difference range $\pi<\varphi<2\pi$. It is shown that the
ABS of $\theta =\pi /4$ and that of $\theta =3\pi /4$ are
symmetric to each other with respect to the $E=0$ axis for the
reason of symmetry. Similarly, the ABS for $\phi =-\pi /4$ is
symmetric to that for $\phi =\pi /4 $ with respect to the $E=0$
axis.

\begin{figure}[tbp]
\begin{center}
\includegraphics[bb=17 15 331 222, width=3.415in]{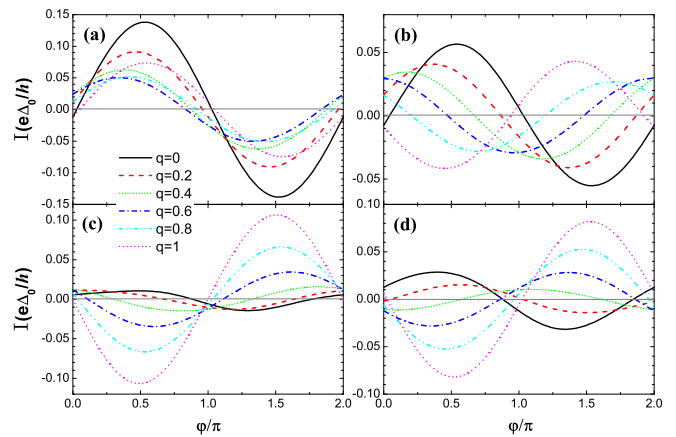}
\end{center}
\caption{(Color online) The CPR for d+p wave with fixed $\protect\theta=%
\protect\pi/4$ and various $\protect\eta$: (a) $\protect\pi/4$, (b) $\protect%
\pi/2$, (c) $3\protect\pi/4$, (d) $\protect\pi$. In each panel, $q$ varies
from $0$ to $1$ with a step of $0.2$. Other parameters are the same as those
in Fig. \protect\ref{sp}.}
\label{dp2}
\end{figure}

The corresponding total Josephson currents are shown in Fig. \ref{dp1}. When
$\theta=0$, the CPR exhibits a $0$-$\pi$ transition with increasing $\eta$.
The anomalous Josephson effect does not occur because the two ABS have
opposite phase shifts as discussed above. When $\theta=\pi/4$, $\theta=\pi/2$%
, and $\theta=3\pi/4$, the anomalous Josephson current appears. The CPR is
the same for $\theta=\pi/4$ and $\theta=3\pi/4$ as the symmetry between
their ABS shows. And the CPR for $\theta=\pi/2$ is only a bit different from
that for $\theta=\pi/4$. As a whole, the ground-state phase difference is
continuously tunable by adjusting $\eta$. When $\theta=\pi/4$ and $\eta=\pi$%
, the contribution from $\phi=\pi/4$ to the supercurrent (not shown here)
show that the propagating state carried supercurrent ($0<\varphi<\pi$, see
Fig. \ref{absdp}) equals to that carried by the ABS ($\pi<\varphi<2\pi$).

Now we discuss the general case of arbitrary $q$ where the two
gaps are both open. We choose $\theta =\pi/4 $ to let the exchange
field align with the spin polarization axis of left-going
(right-going) ABS of $\protect\phi=\protect\pi/4$
($-\protect\pi/4$) component. The CPRs for various $q$ and $\eta $
are shown in Fig. \ref{dp2}. The case of $\eta =0$ is similar to
Fig. \ref{sp} (a) and thus not shown here. For $\eta =\pi /4$ and
$\eta =\pi /2$, the situation is similar to that in the case of
s+p wave. Especially, the case of $\eta =\pi /2$ is still
important to determine the triplet-singlet ratio of NCS because
the ground-state phase difference $\varphi_0$ varies monotonically
with increasing $q$ while the critical current changes little.
What is different from the case of s+p wave is that $\varphi_0$ is
not zero even at $q=0$ or $1$. That is due to the
spin-splitting of the momentum factor of gap function $f(\mathbf{k})$. For $%
\eta =3\pi /4$ and $\eta =\pi$, the evolution of critical current
is similar to that in the $s$+$p$ wave case. But the evolution of
$\varphi_0$ is very different from that for $s$+$p$ wave. Here
$\varphi_0$ changes monotonically with increasing $q$. As
discussed previously, that is because the exchange field is
perpendicular to the spin direction of right-going (left-going)
ABS of $\protect\phi=\protect\pi/4$ ($-\protect\pi/4$) component.
These ABS with perpendicular spin direction makes difference
between singlet and triplet pairing. Thus $\varphi_0$ as well as
the critical current are different for singlet-dominant and
triplet-dominant pairing.

\subsection{$d$+$f$ wave}

For the $d$+$f$ wave case, the momentum dependence of the gap function $f(\mathbf{k}%
)=(k_{x}^{2}-k_{y}^{2})/k_{F}^{2}$. The gap is maximum at $\phi
=0$ or $\pm \pi /2$. However, the contributions from $\phi =\pm
\pi /2$ to the supercurrent are small in comparison with that from
$\phi =0$. The contribution from normal incidence $k_{y}=0$ is
still dominant as in the case of s+p wave. Therefore the situation
for d+f wave is similar to that for s+p wave and then the
numerical results for d+f wave are not presented here.

\section{Conclusion}

In summary, we predicted the appearance of anomalous Josephson
effect with nonzero ground-state phase difference in a NCS/F/NCS
junction. The ground-state phase difference $\varphi_0$ is
proposed to serve as a tool to determine the triplet-singlet ratio
of NCS. The physics picture and analytical results are given on
the basis of $s$+$p$ wave, while the numerical results and discussion
are given on both $s$+$p$ and $d$+$p$ waves. For $s$+$p$ wave, $\varphi_0$
reaches the extremum when the singlet and triplet components have
equal magnitude and there is no difference between
singlet-dominant case and triplet-dominant case generally. But in
the special case of $\eta=\pi/2$, $\varphi_0$ changes
monotonically with increasing singlet percentage $q$. For d+p
wave, the monotonic change of $\varphi_0$ with increasing $q$ is
much more general if only $\eta$ is not too small. Interestingly,
in the case of d+p wave, we also find novel states in which the
supercurrents are totally carried by continuous propagating states
instead of discrete ABS. Instead of carrying supercurrent, the ABS
which here only appear above the Fermi energy block the
supercurrent flowing along the opposite direction. These novel
states advance the understanding of the relation between ABS and
the Josephson current.

\begin{acknowledgments}
The work described in this paper is supported by the National Natural
Science Foundation of China (NSFC, Grant Nos. 11204187, 11204185, and 11274059).
\end{acknowledgments}

\end{document}